\documentclass[11pt,letterpaper]{article}

\usepackage[margin=1in]{geometry}
\usepackage{amsmath}
\usepackage{amsthm}
\usepackage{amsfonts}
\usepackage{amssymb}
\usepackage{mathrsfs}
\usepackage{mathtools} 
\usepackage{xfrac}
\usepackage{booktabs}
\usepackage{stmaryrd} 
\usepackage{authblk}
\newcommand*{\email}[1]{\href{mailto:#1}{\nolinkurl{#1}}}

\usepackage{newtxtext}
\usepackage{lipsum} 
\usepackage[dvipsnames,svgnames]{xcolor}
\usepackage[strict]{changepage} 
\usepackage{framed} 
\usepackage{tcolorbox}
\usepackage{todonotes}
\usepackage{thmtools,thm-restate}
\usepackage[backref=page]{hyperref}
\usepackage[nameinlink]{cleveref}
\usepackage[normalem]{ulem}
\usepackage{braket}
\usepackage{caption}
\usepackage{subcaption}
\usepackage[inline]{enumitem}
\usepackage{algorithm}
\usepackage{algpseudocode}
\usepackage{amsthm}
\newtheorem{definition}{Definition}
\newtheorem{theorem}{Theorem}
\newtheorem{lemma}{Lemma}

\newtheorem{proposition}{Proposition}
\newtheorem{example}{Example}

\usepackage{enumitem}
\setlist[itemize,1]{leftmargin=1em}
\setlist[itemize,2]{leftmargin=1em}
\setlist[itemize,3]{leftmargin=1em}
\tcbuselibrary{most}

\newcommand{\RR}{\mathbb{R}}

\newcommand{\QQ}{\mathbb{Q}}

\newcommand{\eps}{\varepsilon}

\DeclareMathOperator*{\argmax}{arg\,max}
\DeclareMathOperator{\supp}{supp}
\DeclareMathOperator{\relint}{relint}
\DeclareMathOperator{\conv}{conv}

\def\np{\textup{\textsf{NP}}}

\def\ch{\textup{\textsf{CH}}}
\def\pspace{\textup{\textsf{PSPACE}}}

\def\PPAD{\textup{\textsf{PPAD}}}
\def\FIXP{\textup{\textsf{FIXP}}}
\def\SDFP/{\textup{\textsf{SDFP}}}

\def\SAT{\textup{\textsf{SAT}}}

\begin{document}

\title{On Existence and Computational Complexity of\\Quasi-strict Equilibria}

\author[*]{Kristoffer Arnsfelt Hansen}
\affil[*]{Aarhus University\\ \email{arnsfelt@cs.au.dk}}
\author[$\dagger$]{Xinhao Nie}
\affil[$\dagger$]{Aarhus University\\ \email{nie@cs.au.dk}}

\date{}

\maketitle
\thispagestyle{empty}

\begin{abstract}
Quasi-strict Nash equilibrium, introduced by Harsanyi in 1973, is the Nash equilibrium refinement that requires every pure best response to be played with positive probability. Norde established that any bimatrix game admits a quasi-strict Nash equilibrium. We give a substantially simpler proof of this result that in particular avoids an intricate limit argument used in the final step of the proof by Norde.

Our existence proof allows us to positively resolve the outstanding open question of $\PPAD$ membership of the problem of computing a quasi-strict Nash equilibrium, thereby settling its precise complexity. In games with three or more players, a quasi-strict Nash equilibrium is not guaranteed to exist, leading to the natural problem of deciding the existence of a quasi-strict Nash equilibrium in a given strategic form game. We show that this problem is $\exists\RR$-complete, strengthening a previous $\np$-hardness result.
\end{abstract}

\section{Introduction}
The fundamental theorem of Nash establishes that every finite strategic form game admits a Nash equilibrium in mixed strategies~\cite{nash1951noncooperative}, i.e.~a mixed-strategy profile in which every player's strategy is a best response to the strategies of the other players. Nash equilibrium \emph{refinements} impose additional requirements with the aim of capturing certain desirable properties. In this work we are concerned with the notion of \emph{quasi-strict} Nash equilibria introduced by Harsanyi~\cite{harsanyi1973games}, which simply requires each player to play each pure best response to the strategies of the other players with positive probability. In subsequent work Harsanyi~\cite[Theorem~2]{harsanyi1973oddness} proved that in almost all finite strategic form games all Nash equilibria are in fact quasi-strict equilibria. In general, however, quasi-strict equilibria are not guaranteed to exist in games with~3 or more players~\cite[Fig~3.4.1]{vanDamme1991stability} and when they do exist they may necessarily assign positive probability to dominated pure strategies even in bimatrix games~\cite[Fig~2.1.1]{vanDamme1991stability}. On the other hand, as shown by Squires~\cite{squires1998impossibility}, from any quasi-strict equilibrium one can construct sets satisfying the axioms introduced by Cubitt and Sugden~\cite{cubitt1994rationally} as formal conditions for justifiable play, making them very interesting from an axiomatic point of view.

Norde proved that quasi-strict equilibria exist in \emph{all} bimatrix games~\cite{norde1999bimatrix}. A fundamental technical challenge in proving this result is that it is not possible to express the set of quasi-strict equilibria as the set of fixed points given by the fixed point theorems of Brouwer~\cite{brouwer1911ueber} and Kakutani~\cite{kakutani1941generalization}, simply because these are closed sets which the set of quasi-strict equilibria may fail to be.
This does of course not rule out expressing a closed subset of the quasi-strict equilibria as a set of fixed points, but it is not clear how to do that in a general way. Norde overcomes the obstacles presented by a direct fixed point formulation by several means. First, for each set $T$ of pure strategies of player~2, a generalized permutahedron $K_T$ is constructed in the relative interior of the face of the strategy simplex $\Delta_m$ of player~2 defined by $T$, chosen to approximate this relative interior sufficiently well. On each such set $K_T$, a continuous best-response function $f_T \colon K_T \to \Delta_n$ for player~1 is constructed and a corresponding best-response function $g^k_T \colon \Delta_n \to K_T$ for player~2 defined using projection and a scaling of the payoffs by a large positive constant~$k$. A continuous function $h_k \colon \Delta_m \to \Delta_m$ is then defined as a weighted combination of the functions $f_T \circ g_T^k$ to which Brouwer's fixed point theorem applies. Finally, it is proved that limit points of fixed points of the functions $h_k$, as $k$ tends to infinity, can form the strategy of player~1 in a quasi-strict equilibrium.

In this work we are interested in quasi-strict equilibria from the viewpoint of computational complexity. The seminal work of Papadimitriou~\cite{papadimitriou1994complexity} introduced the complexity class $\PPAD$ to capture total search problems whose totality is given by directed parity arguments.
These include, in particular, search problems arising from path-following procedures underlying  computational versions of Sperner's lemma and Lemke's algorithm. Using these, Papadimitriou showed that the problems of computing a Nash equilibrium of bimatrix games and of computing an $\eps$-approximate Nash equilibrium in a strategic form game belong to $\PPAD$. 
Later, the celebrated works of Daskalakis, Goldberg, and Papadimitriou~\cite{daskalakis2009complexity} and Chen, Deng, and Teng~\cite{chen2009settling} proved that these problems are in fact $\PPAD$-complete.

The $\PPAD$-hardness of computing a Nash equilibrium of a bimatrix game trivially implies $\PPAD$-hardness of computing a quasi-strict equilibrium in bimatrix games. Brandt and Fischer~\cite{brandt2008hardness} gave a polynomial time algorithm for finding a quasi-strict equilibrium in two-player zero-sum games and symmetric multi-player games in which every player has exactly~two pure strategies. The precise complexity of computing a quasi-strict equilibrium in bimatrix games remained an outstanding open problem. Our main result is settling this question by proving that the problem belongs to $\PPAD$.
\begin{restatable}{theorem}{ppad}
    \label{thm:qsne-ppad}
    Computing a quasi-strict Nash equilibrium of a bimatrix game is $\PPAD$-complete.
\end{restatable}
On the way to proving our result we first develop an alternative proof of Norde's existence theorem. This proof relies primarily on simplicial complexes and simplicial maps and allows us to bypass the technically involved arguments employed in Norde's proof. Our proof also follows a more natural structure of treating each player symmetrically. More specifically, for each player we consider the polyhedral complex defined by the best-response regions and refine it to a simplicial complex by taking geometric realizations of the order complexes of its face posets. In addition to this, we also consider the barycentric subdivisions of each strategy simplex, giving for each player two simplicial complexes with the same underlying space. On these we define natural simplicial maps defined in terms of the support of the strategy defined by the vertex and the set of pure best replies to the strategy defined by the vertex. These are combined to form a single continuous self-map $T$ on the set of mixed strategy profiles using radial projection of a translation of the given strategy profile in a
way similar to the existence proof of Nash equilibria by Nash~\cite{nash1951noncooperative}. Finally, by examining the inclusion relations among faces, we show how to extract a quasi-strict Nash equilibrium directly from a fixed point of~$T$.

To prove our $\PPAD$-membership result we make use of the notion of polynomial piecewise linear fixed point problems, as defined and used by Etessami and Yannakakis~\cite{etessami2010complexity} in their characterization of $\PPAD$ by  the search problems given by the piecewise linear fragment of the general complexity class $\FIXP$. An instance $I$ of a polynomial piecewise linear fixed point problem  is given by a continuous self-map $F_I$ of a nonempty compact convex domain $D_I$,  divided in polyhedral cells on which $F_I$ is linear. The domain $D_I$ is given by a set of linear inequalities with rational coefficients computable from $I$ and the linear function, which must be defined by a rational matrix and vector, of a given cell containing a point $x$ must be computable in polynomial time by an oracle algorithm that can perform linear comparison queries of the form $a^\top x\leq b$ with rational-valued $a$ and $b$.

To use this notion requires us to make our existence proof effective. In particular, for refining the polyhedral complex defined by the best-response regions we define the interior points chosen of each face implicitly by solutions of linear programs. This allows them to be computed in polynomial time whenever they are needed, which in turn allows us to implement the necessary oracle algorithm in polynomial time. To make the self-map piecewise linear we also replace the radial projection by Euclidean projection. The same change of the type of projection was previously used by Etessami and Yannakakis~\cite{etessami2010complexity} in their characterization of the set of Nash equilibria as fixed points of continuous functions defined without use of division.

Our second result is concerned with games with three or more players, where a quasi-strict Nash equilibrium is not guaranteed to exist. This naturally leads to the decision problem of deciding existence of a quasi-strict Nash equilibrium in a given strategic form game. From the definition of quasi-strict equilibria it is immediate to see that this problem belongs to the complexity class~$\exists\RR$ which is the complexity class associated with the decision problem for the existential theory of the reals. Brandt and Fischer~\cite{brandt2008hardness} proved that the problem is $\np$-hard by adapting a construction of McLennan and Tourky~\cite{mclennan2010simplecomplexity} used for proving $\np$-completeness for other decision problems about Nash equilibria. We show that the problem is in fact $\exists\RR$-hard, thereby settling the complexity of the problem.
\begin{restatable}{theorem}{existr}
    \label{thm:qsne-existence-etr}
    For every fixed $n\ge3$, deciding whether an $n$-player game
    in strategic form with rational payoffs admits a quasi-strict
    Nash equilibrium is $\exists\RR$-complete, even for zero-sum games.
\end{restatable}
The proof is based on $\exists\RR$-hardness of deciding whether the minmax value of player~1 in a 3-player game is at most~0, which, importantly for our purposes, holds even when a certain useful property of the payoff functions is satisfied, which in particular enables player~1 to ensure that the minmax value is at least~0,
and under a promise that a minmax value of~0, when possible, may be ensured by fully mixed strategies of player~2 and player~3~\cite{hansen2019real}. We follow the definition of a corresponding zero-sum game used by Berthelsen and Hansen~\cite{berthelsen2022computational} to prove $\exists\RR$-completeness for other decision problems about Nash equilibria in multi-player zero-sum games. Similarly to proofs of these results, our reduction simply gives each player a new strategy in this game and we define payoffs that ensure that the resulting game has a quasi-strict equilibrium if and only if the minmax value of player~1 in the original game is~0.

\subsection{Related work}
A substantial body of research studies the computational complexity of computing various notions of Nash equilibrium refinements. We shall here focus on strategic form games. Nash equilibrium refinements that are guaranteed to exist naturally inherit the computational hardness of computing \emph{any} Nash equilibrium, i.e.\ the $\PPAD$-completeness of computing a Nash equilibrium in bimatrix games~\cite{daskalakis2009complexity,chen2009settling} and the $\FIXP$-completeness of computing a Nash equilibrium in multiplayer games~\cite{etessami2010complexity}. 

Classic equilibrium refinements defined to take the possibility of players making \emph{mistakes} into account are the notions of \emph{perfect} equilibria due to Selten~\cite{selten1975reexamination} and \emph{proper} equilibria due to Myerson~\cite{myerson1978refinements}. Perfect equilibria in two-player games coincide with a Nash equilibrium wherein neither strategy is dominated. This means that a perfect equilibrium in two-player zero-sum games may be found in polynomial time using linear programming. For bimatrix games computing a perfect equilibrium  belongs to $\PPAD$, as a consequence of the complementary pivoting algorithm for computing perfect equilibria in two-player extensive form games by von Stengel, van den Elzen, and Talman~\cite{vonStengel2002computing}. For multiplayer games Etessami, Hansen, Miltersen and Sørensen~\cite{etessami2014trembling} showed that the problem belongs to $\FIXP_a$, i.e.\ that the problem of computing a strategy profile that is $\eps$-close to a perfect equilibrium for a given $\eps>0$ belongs to $\FIXP$. All these results were subsequently strengthened to the same results for proper equilibria~\cite{myerson1978refinements}. A polynomial time algorithm for zero-sum games was given by Miltersen and Sørensen~\cite{Miltersen2006proper}, Sørensen~\cite{sorensen2012proper} proved the problem to be in $\PPAD$ for bimatrix games, and Hansen and Lund~\cite{hansen2018proper} proved the problem to belong  to $\FIXP_a$ for multiplayer games. Thus for two-player games, the complexity of finding a perfect or proper equilibrium is the same as the complexity of finding a quasi-strict equilibrium. This is not the case for games with three or more  players, simply because a quasi-strict equilibrium may fail to exist.

Another interesting Nash equilibrium refinement is the notion of \emph{test-set} equilibrium due to Milgrom and Mollner~\cite{milgrom2018equilibriumselection}. For two-player games, every proper equilibrium is a test-set equilibrium, and every test-set equilibrium is a perfect equilibrium~\cite[Proposition~1]{milgrom2018equilibriumselection}, which means that computing a test-set equilibrium can be done in polynomial time for two-player zero-sum games and belongs to $\PPAD$ for bimatrix games. But, similarly to quasi-strict equilibria, a test-set equilibrium may fail to exist in games with three or more players~\cite[Figure~1]{milgrom2018equilibriumselection}.

\section{Preliminaries}

For our existence proof we make use of fundamental topological notions, where we generally follow the terminology of Munkres~\cite{munkres2000topology, munkres2018elements}. For polyhedral complexes we adopt the definition of Björner~\cite{bjorner1995topological}. The concepts
and results we need have natural geometric interpretations, which we illustrate by example using a bimatrix game.

For $k\geq 1$, we let $[k]$ denote the set $\{1,\ldots,k\}$ and let $\Delta_k$ denote the standard $(k-1)$-simplex given by
\[
    \Delta_k
    \coloneq 
    \left\{
        z\in\RR^k:
        z_i\geq 0\text{ for all }i\in[k],\
        \sum_{i=1}^k z_i=1
    \right\}.
\]
A point $z\in\Delta_k$ can be viewed as a probability distribution on $[k]$ and we define $\supp(z)=\{i\in[k]:z_i>0\}$.

\begin{definition}[Two-player Bimatrix games and Quasi-strict Nash Equilibria]
    \label{def:quasi-strict nash eq}
    Let $M \coloneq [m]$ be the set of pure strategies of player~1 and $N \coloneq [n]$ be the set of pure strategies of player~2. We let $x \in \Delta_m$ and $y \in \Delta_n$ denote the mixed strategies of player~1 and player~2, respectively. A bimatrix game is given by a pair $I \coloneq (A,B)$ of matrices $A,B \in \RR^{m \times n}$ specifying payoffs  of player~1 and player~2, respectively.
    For the matrix~$A$ and probability vector~$y$, the set
    \[
    BR_1(y) \coloneqq \{i \in M : i \in \argmax_{j \in M} e_j^\top Ay \}
    \]
    is the set of pure best responses of player~1 to strategy~$y$. For the matrix~$B$ and
    probability vector~$x$, the set
    \[
    BR_2(x) \coloneqq \{i \in N: i \in \argmax_{j \in N} x^\top B e_j \}
    \]
    is the set of pure best responses of player~2 to strategy~$x$.

    A Nash equilibrium is a pair of mixed strategies $(x^*, y^*)$ satisfying
    \[
    \supp(x^*) \subseteq BR_1(y^*) \qquad \text{and} \qquad \supp(y^*) \subseteq BR_2(x^*).
    \]
    A quasi-strict Nash equilibrium further requires that 
    \[
    \supp(x^*) = BR_1(y^*) \qquad \text{and} \qquad \supp(y^*) = BR_2(x^*).
    \]
\end{definition}

When considering the computational problem of finding an equilibrium, the payoff matrices are defined on the rational number field $\QQ$.
The following example is adapted from \cite{norde1999bimatrix}, with a 
modified payoff matrix for player~2.
For clarity, we denote player~1's pure action $i$ by $x_i$ and
player~2's pure action $j$ by $y_j$, rather than using the notation $e_i$.
\begin{example}
    \label{example:bimatrix game}
    Consider the bimatrix game with payoff matrices
    \[
        A =
        \begin{pmatrix}
            -1 & 2 & 1 \\
             1 & 0 & 0 \\
             0 & 1 & 1
        \end{pmatrix},
        \qquad
        B =
        \begin{pmatrix}
            1 & 0 & 0 \\
            0 & 1 & 0 \\
            0 & 0 & 1
        \end{pmatrix}.
    \]
    The best-response relations $BR_1$ and $BR_2$ subdivide the strategy simplices
    $\Delta_n$ and $\Delta_m$, respectively.
    As illustrated by the calculation, each closed best-response region is a polytope.
    For example,
    write $y=q_1y_1+q_2y_2+q_3y_3$, where $q_i\ge0$ and
    $q_1+q_2+q_3=1$. Player~1's pure action~$x_1$ is the unique best
    response precisely when
    \[
    \begin{aligned}
        BR_1(y)=\{1\}
        &\iff
        \begin{cases}
            -q_1+2q_2+q_3>q_1,\\
            -q_1+2q_2+q_3>q_2+q_3
        \end{cases}.
    \end{aligned}
    \]
    Then, $q_2>q_1$ and $q_1+q_2\le1$ imply $q_1<\tfrac12$.
    For $x=p_1x_1+p_2x_2+p_3x_3$, where $p_i\ge0$ and
    $p_1+p_2+p_3=1$, the choice $B=I_3$ gives
    \[
        BR_2(x)=\argmax_{j\in[3]} \ p_j.
    \]
    \begin{center}
\captionsetup{type=figure,font=small}
\resizebox{0.75\linewidth}{!}{%
\begin{tikzpicture}[x=1cm,y=1cm,font=\normalfont\small,line width=.55pt]
    \definecolor{brone}{HTML}{2761A3}
    \definecolor{brtwo}{HTML}{B6601C}
    \definecolor{brthree}{HTML}{32734A}
    \begin{scope}
        \coordinate (Y1) at (0,0);
        \coordinate (Y2) at (6,0);
        \coordinate (Y3) at (3,5.1961524);
        \coordinate (M) at (3,0);
        \coordinate (N) at (1.5,2.5980762);

        \draw (Y1)--(Y2)--(Y3)--cycle;
        \draw (Y3)--(M) (N)--(M);

        \node[text=brone] at (4.15,1.65) {$1$};
        \node[text=brtwo] at (1.35,.78) {$2$};
        \node[text=brthree] at (2.38,3.00) {$3$};

        \foreach \vertex in {M,N}
            \fill (\vertex) circle (2pt);
        \node[below=4pt] at (Y1) {$y_1$};
        \node[below=4pt] at (Y2) {$y_2$};
        \node[above=4pt] at (Y3) {$y_3$};
        \node[below=4pt] at (M) {$\tfrac12 y_1+\tfrac12 y_2$};
        \node[left=4pt] at (N) {$\tfrac12 y_1+\tfrac12 y_3$};
        \node at (3,5.95) {$BR_1$ on $\Delta_n$};
    \end{scope}

    \begin{scope}[xshift=7.75cm]
        \coordinate (X1) at (0,0);
        \coordinate (X2) at (6,0);
        \coordinate (X3) at (3,5.1961524);
        \coordinate (M12) at (3,0);
        \coordinate (M13) at (1.5,2.5980762);
        \coordinate (M23) at (4.5,2.5980762);
        \coordinate (C) at (3,1.7320508);

        \draw (X1)--(X2)--(X3)--cycle;
        \draw (M12)--(C) (M13)--(C) (M23)--(C);

        \node[text=brone] at (1.6,.90) {$1$};
        \node[text=brtwo] at (4.4,.90) {$2$};
        \node[text=brthree] at (3,3.45) {$3$};

        \foreach \vertex in {M12,M13,M23,C}
            \fill (\vertex) circle (2pt);
        \node[below=4pt] at (X1) {$x_1$};
        \node[below=4pt] at (X2) {$x_2$};
        \node[above=4pt] at (X3) {$x_3$};
        \node[below=4pt] at (M12) {$\tfrac12 x_1+\tfrac12 x_2$};
        \node[left=4pt] at (M13) {$\tfrac12 x_1+\tfrac12 x_3$};
        \node[right=4pt] at (M23) {$\tfrac12 x_2+\tfrac12 x_3$};
        \node[below=2pt] at (C) {$\tfrac13 x_1+\tfrac13 x_2+\tfrac13 x_3$};
        \node at (3,5.95) {$BR_2$ on $\Delta_m$};
    \end{scope}
\end{tikzpicture}%
}
\caption{Best-response partitions induced by $BR_1$ on $\Delta_n$ (left)
and $BR_2$ on $\Delta_m$ (right).}
\label{fig:best-response-partitions}
\end{center}

\end{example}

\begin{definition}[Simplex and Simplicial Complex]
    Let $\{a_0,a_1,\dots,a_n\}$ be an affinely independent point set in $\RR^n$.
    The $n$-simplex spanned by these points is the set of all points $x \in \RR^n$ such that
    \[
    x = \sum_{i=0}^n t_i a_i
    \]
    where $\sum_{i=0}^n t_i = 1$ and $t_i \geq 0$
    for each $i$. For each point $x$ in the simplex, the coefficients $t_i$ are uniquely determined.
    They are called the \emph{barycentric coordinates} of $x$ with respect to $a_0,\ldots,a_n$.
    A simplicial complex $\mathcal K$ in
    $\RR^n$ is a collection of simplices in $\RR^n$ such that:
    \begin{enumerate}
    \item Every face of a simplex of $\mathcal K$ is in $\mathcal K$.
    \item The intersection of any two simplices of $\mathcal K$ is a face of each of them.
    \end{enumerate}
    
\end{definition}

\begin{example}
In \Cref{example:bimatrix game}, the best-response regions on $\Delta_n$,
together with all their faces, form a simplicial complex.
On $\Delta_m$, the best-response regions are quadrilaterals, so their
subdivision is not simplicial.
\end{example}

Thus, the best-response subdivisions are in general \emph{not}
simplicial complexes. To facilitate the construction of a continuous self-map,
we use topological tools to obtain a finer subdivision that is a simplicial complex.
This construction is described in \cite[\S 9.3, \S 12.2]{bjorner1995topological}
and also appears as an exercise in
\cite[\S 3, Ex.~4]{munkres2018elements}.

\begin{definition}[Polytopes and polyhedral complexes]
    A convex polytope $P$ is a bounded subset of $\RR^d$
    which is the solution set of a finite number of linear
    equalities and inequalities.
    Any nonempty subset obtained by changing some of the
    inequalities to equalities is a face of $P$.

    A polyhedral complex 
    $\mathcal C$ is a finite collection of convex polytopes
    in $\RR^d$ such that:
    \begin{enumerate}
        \item Every face of a polytope of $\mathcal C$ is in $\mathcal C$.

        \item The intersection of any two polytopes of $\mathcal C$ is a face of each of them.
    \end{enumerate}
\end{definition}

\begin{example}
    In \Cref{example:bimatrix game}, the closed
    best-response regions on $\Delta_m$, together
    with all their nonempty faces, form a polyhedral
    complex. The best-response regions on $\Delta_n$
    could be also considered as a polyhedral complex. 
\end{example} 

\begin{definition}[Face poset and order complex]
    Let $\mathcal C$ be a polyhedral complex.
    Its face poset
    \[
        \mathcal P(\mathcal C)
        =
        (\mathcal C,\subseteq)
    \]
    consists of its faces ordered by inclusion.

    For a finite poset $(\mathcal P,\preceq)$, the
    order complex is the
    abstract simplicial complex on vertex set
    $\mathcal P$ whose $r$-simplices are the
    $(r+1)$-element chains
    \[
        F_0\prec F_1\prec\cdots\prec F_r.
    \]
    In particular, the simplices 
    correspond to chains of faces
    \[
    F_0\subsetneq F_1\subsetneq\cdots\subsetneq F_r,
    \qquad r\geq 0.
    \]
\end{definition}

\begin{example}
    In \Cref{example:bimatrix game}, each maximal chain
    in the face poset has the form
    \[
        \{v\}\subsetneq e\subsetneq P,
    \]
    consisting of a vertex, an incident edge, and a
    two-dimensional polytope.
    It defines a 2-simplex in the
    corresponding order complex.
\end{example}

Write $\relint F$ for the interior of $F$ in its
affine hull.
For each face $F\in\mathcal C$, fix a single point
\[
    p_F\in\relint F.
\]
The same selected point $p_F$ is used whenever $F$
occurs in a chain.
Set
\[
    \mathcal K(\mathcal C,p)
    =
    \left\{
        \conv\{p_{F_0},\ldots,p_{F_r}\}:
        F_0\subsetneq\cdots\subsetneq F_r,\
        F_i\in\mathcal C,\ r\geq 0
    \right\}.
\]

We use the following standard results without proof; see 
\cite[\S 12.2]{bjorner1995topological}.

\begin{proposition}
    \label{prop:subdivision}
    The collection $\mathcal K(\mathcal C,p)$ is
    a simplicial complex with the same underlying
    space as $\mathcal C$:
    \[
        |\mathcal K(\mathcal C,p)|=|\mathcal C|.
    \]
\end{proposition}

\begin{example}
    \label{example: barycenter of best-response}
    In \Cref{example:bimatrix game}, let
    \[
        Q_i
        =
        \{y\in\Delta_n:i\in\operatorname{BR}_1(y)\},
        \qquad
        P_j
        =
        \{x\in\Delta_m:j\in\operatorname{BR}_2(x)\}
    \]
    denote the closed best-response regions.
    Let $\mathcal C_n$ and $\mathcal C_m$ consist of all
    their nonempty faces, respectively.

    For each face $F$, take $b_F$ to be the average
    of its vertices and realize each chain by
    \[
        F_0\subsetneq\cdots\subsetneq F_r
        \quad\longmapsto\quad
        \operatorname{conv}\{b_{F_0},\ldots,b_{F_r}\}.
    \]
    This gives barycentric subdivisions of the
    best-response complexes.
    By \Cref{prop:subdivision},
    their underlying spaces are $\Delta_n$
    and $\Delta_m$, respectively.

    \begin{center}
\captionsetup{type=figure,font=small}
\resizebox{0.75\linewidth}{!}{%
\begin{tikzpicture}[x=1cm,y=1cm,font=\normalfont\small,line width=.55pt]
    \definecolor{brone}{HTML}{2761A3}
    \definecolor{brtwo}{HTML}{B6601C}
    \definecolor{brthree}{HTML}{32734A}
    \begin{scope}[xshift={0 cm}]
        \fill[brone!20] (6,0)--(3,5.19615242)--(3,0)--cycle;
        \fill[brtwo!20] (0,0)--(3,0)--(1.5,2.59807621)--cycle;
        \fill[brthree!20] (3,5.19615242)--(1.5,2.59807621)--(3,0)--cycle;
        \draw[line width=.35pt] (4.00000000,1.73205081)--(6.00000000,0.00000000);
        \draw[line width=.35pt] (4.00000000,1.73205081)--(3.00000000,5.19615242);
        \draw[line width=.35pt] (4.00000000,1.73205081)--(3.00000000,0.00000000);
        \draw[line width=.35pt] (4.00000000,1.73205081)--(4.50000000,2.59807621);
        \draw[line width=.35pt] (4.00000000,1.73205081)--(3.00000000,2.59807621);
        \draw[line width=.35pt] (4.00000000,1.73205081)--(4.50000000,0.00000000);
        \draw[line width=.35pt] (1.50000000,0.86602540)--(0.00000000,0.00000000);
        \draw[line width=.35pt] (1.50000000,0.86602540)--(3.00000000,0.00000000);
        \draw[line width=.35pt] (1.50000000,0.86602540)--(1.50000000,2.59807621);
        \draw[line width=.35pt] (1.50000000,0.86602540)--(1.50000000,0.00000000);
        \draw[line width=.35pt] (1.50000000,0.86602540)--(2.25000000,1.29903811);
        \draw[line width=.35pt] (1.50000000,0.86602540)--(0.75000000,1.29903811);
        \draw[line width=.35pt] (2.50000000,2.59807621)--(3.00000000,5.19615242);
        \draw[line width=.35pt] (2.50000000,2.59807621)--(1.50000000,2.59807621);
        \draw[line width=.35pt] (2.50000000,2.59807621)--(3.00000000,0.00000000);
        \draw[line width=.35pt] (2.50000000,2.59807621)--(2.25000000,3.89711432);
        \draw[line width=.35pt] (2.50000000,2.59807621)--(2.25000000,1.29903811);
        \draw[line width=.35pt] (2.50000000,2.59807621)--(3.00000000,2.59807621);
        \draw (6.00000000,0.00000000)--(3.00000000,5.19615242);
        \draw (3.00000000,5.19615242)--(3.00000000,0.00000000);
        \draw (6.00000000,0.00000000)--(3.00000000,0.00000000);
        \draw (0.00000000,0.00000000)--(3.00000000,0.00000000);
        \draw (3.00000000,0.00000000)--(1.50000000,2.59807621);
        \draw (0.00000000,0.00000000)--(1.50000000,2.59807621);
        \draw (3.00000000,5.19615242)--(1.50000000,2.59807621);
        \fill (0.00000000,0.00000000) circle (2pt);
        \fill (6.00000000,0.00000000) circle (2pt);
        \fill (3.00000000,5.19615242) circle (2pt);
        \fill (3.00000000,0.00000000) circle (2pt);
        \fill (1.50000000,2.59807621) circle (2pt);
        \fill (4.50000000,2.59807621) circle (2pt);
        \fill (3.00000000,2.59807621) circle (2pt);
        \fill (4.50000000,0.00000000) circle (2pt);
        \fill (1.50000000,0.00000000) circle (2pt);
        \fill (2.25000000,1.29903811) circle (2pt);
        \fill (0.75000000,1.29903811) circle (2pt);
        \fill (2.25000000,3.89711432) circle (2pt);
        \fill (4.00000000,1.73205081) circle (2pt);
        \fill (1.50000000,0.86602540) circle (2pt);
        \fill (2.50000000,2.59807621) circle (2pt);
        \node[below left=3pt] at (0.00000000,0.00000000) {$y_1$};
        \node[below right=3pt] at (6.00000000,0.00000000) {$y_2$};
        \node[above=4pt] at (3.00000000,5.19615242) {$y_3$};
        \node at (3,5.95) {$BR_1$ on $\Delta_n$};
    \end{scope}
    \begin{scope}[xshift={7.75 cm}]
        \fill[brone!20] (0,0)--(3,0)--(3,1.73205081)--(1.5,2.59807621)--cycle;
        \fill[brtwo!20] (6,0)--(4.5,2.59807621)--(3,1.73205081)--(3,0)--cycle;
        \fill[brthree!20] (3,5.19615242)--(1.5,2.59807621)--(3,1.73205081)--(4.5,2.59807621)--cycle;
        \draw[line width=.35pt] (1.87500000,1.08253175)--(0.00000000,0.00000000);
        \draw[line width=.35pt] (1.87500000,1.08253175)--(3.00000000,0.00000000);
        \draw[line width=.35pt] (1.87500000,1.08253175)--(3.00000000,1.73205081);
        \draw[line width=.35pt] (1.87500000,1.08253175)--(1.50000000,2.59807621);
        \draw[line width=.35pt] (1.87500000,1.08253175)--(1.50000000,0.00000000);
        \draw[line width=.35pt] (1.87500000,1.08253175)--(3.00000000,0.86602540);
        \draw[line width=.35pt] (1.87500000,1.08253175)--(2.25000000,2.16506351);
        \draw[line width=.35pt] (1.87500000,1.08253175)--(0.75000000,1.29903811);
        \draw[line width=.35pt] (4.12500000,1.08253175)--(6.00000000,0.00000000);
        \draw[line width=.35pt] (4.12500000,1.08253175)--(4.50000000,2.59807621);
        \draw[line width=.35pt] (4.12500000,1.08253175)--(3.00000000,1.73205081);
        \draw[line width=.35pt] (4.12500000,1.08253175)--(3.00000000,0.00000000);
        \draw[line width=.35pt] (4.12500000,1.08253175)--(5.25000000,1.29903811);
        \draw[line width=.35pt] (4.12500000,1.08253175)--(3.75000000,2.16506351);
        \draw[line width=.35pt] (4.12500000,1.08253175)--(3.00000000,0.86602540);
        \draw[line width=.35pt] (4.12500000,1.08253175)--(4.50000000,0.00000000);
        \draw[line width=.35pt] (3.00000000,3.03108891)--(3.00000000,5.19615242);
        \draw[line width=.35pt] (3.00000000,3.03108891)--(1.50000000,2.59807621);
        \draw[line width=.35pt] (3.00000000,3.03108891)--(3.00000000,1.73205081);
        \draw[line width=.35pt] (3.00000000,3.03108891)--(4.50000000,2.59807621);
        \draw[line width=.35pt] (3.00000000,3.03108891)--(2.25000000,3.89711432);
        \draw[line width=.35pt] (3.00000000,3.03108891)--(2.25000000,2.16506351);
        \draw[line width=.35pt] (3.00000000,3.03108891)--(3.75000000,2.16506351);
        \draw[line width=.35pt] (3.00000000,3.03108891)--(3.75000000,3.89711432);
        \draw (0.00000000,0.00000000)--(3.00000000,0.00000000);
        \draw (3.00000000,0.00000000)--(3.00000000,1.73205081);
        \draw (1.50000000,2.59807621)--(3.00000000,1.73205081);
        \draw (0.00000000,0.00000000)--(1.50000000,2.59807621);
        \draw (6.00000000,0.00000000)--(4.50000000,2.59807621);
        \draw (4.50000000,2.59807621)--(3.00000000,1.73205081);
        \draw (3.00000000,0.00000000)--(6.00000000,0.00000000);
        \draw (1.50000000,2.59807621)--(3.00000000,5.19615242);
        \draw (4.50000000,2.59807621)--(3.00000000,5.19615242);
        \fill (0.00000000,0.00000000) circle (2pt);
        \fill (6.00000000,0.00000000) circle (2pt);
        \fill (3.00000000,5.19615242) circle (2pt);
        \fill (3.00000000,0.00000000) circle (2pt);
        \fill (1.50000000,2.59807621) circle (2pt);
        \fill (4.50000000,2.59807621) circle (2pt);
        \fill (3.00000000,1.73205081) circle (2pt);
        \fill (1.50000000,0.00000000) circle (2pt);
        \fill (3.00000000,0.86602540) circle (2pt);
        \fill (2.25000000,2.16506351) circle (2pt);
        \fill (0.75000000,1.29903811) circle (2pt);
        \fill (5.25000000,1.29903811) circle (2pt);
        \fill (3.75000000,2.16506351) circle (2pt);
        \fill (4.50000000,0.00000000) circle (2pt);
        \fill (2.25000000,3.89711432) circle (2pt);
        \fill (3.75000000,3.89711432) circle (2pt);
        \fill (1.87500000,1.08253175) circle (2pt);
        \fill (4.12500000,1.08253175) circle (2pt);
        \fill (3.00000000,3.03108891) circle (2pt);
        \node[below left=3pt] at (0.00000000,0.00000000) {$x_1$};
        \node[below right=3pt] at (6.00000000,0.00000000) {$x_2$};
        \node[above=4pt] at (3.00000000,5.19615242) {$x_3$};
        \node at (3,5.95) {$BR_2$ on $\Delta_m$};
    \end{scope}
\end{tikzpicture}%
}
\caption{Barycentric subdivisions of the best-response complexes on
$\Delta_n$ (left) and $\Delta_m$ (right).
Colors indicate the original best-response regions.}
\label{fig:barycentric-realization}
\end{center}

\end{example}

\begin{definition}[Simplicial map]
    Let $K$ and $L$ be simplicial complexes, and let
    \[
        f:K^{(0)}\to L^{(0)}
    \]
    be a map.
    Suppose that whenever $v_0,\ldots,v_r$ span
    a simplex of $K$, their images
    $f(v_0),\ldots,f(v_r)$ are vertices of
    a simplex of $L$.

    Define $g:|K|\to|L|$ on each simplex by
    \[
        g\left(\sum_{i=0}^r t_i v_i\right)
        =
        \sum_{i=0}^r t_i f(v_i),
        \qquad
        t_i\geq 0,\qquad
        \sum_{i=0}^r t_i=1.
    \]
    The map $g$ is called the simplicial map
    induced by the vertex map $f$.
\end{definition}

We have the following fundamental property of simplicial maps.
\begin{proposition}
    \label{prop:simplicial map}
    The simplicial map $g:|K|\to|L|$ induced by a vertex map satisfying 
    the above condition is well-defined and continuous.            
\end{proposition}

\begin{example}
    For the game in \Cref{example:bimatrix game}, where $m=n=3$,
    give the target strategy simplex $\Delta_3$ its barycentric
    subdivision. On the vertices $b_F$ of the source subdivision
    in \Cref{example: barycenter of best-response}, define
    \[
        R_1(b_F)=
        \frac{1}{|BR_1(b_F)|}
        \sum_{i\in BR_1(b_F)}x_i,
        \qquad F\in\mathcal C_n.
    \]
    Thus each vertex is sent to the barycenter of the face
    spanned by its pure best responses.
    Along a face chain, the best-response sets are nested
    in reverse order, so the image vertices span a simplex
    in the target subdivision. Extending affinely over each
    source simplex gives a simplicial map
    $R_1:\Delta_3\to\Delta_3$.

    Let $M=(y_1+y_2)/2$, and let $q_0,q_1,q_2$ be the
    barycenters of the faces in the chain
    \[
        \{M\}\subsetneq\conv\{M,y_3\}\subsetneq Q_1.
    \]
    Their best-response sets are $\{1,2,3\}$, $\{1,3\}$,
    and $\{1\}$, respectively, so the vertex rule gives
    \[
        r_0=R_1(q_0)=\frac{x_1+x_2+x_3}{3},
        \qquad
        r_1=R_1(q_1)=\frac{x_1+x_3}{2},
        \qquad
        r_2=R_1(q_2)=x_1.
    \]
    Extending $q_i\mapsto r_i$ affinely maps the shaded
    simplex to the shaded target simplex.
    \begin{center}
\captionsetup{type=figure,font=small}
\resizebox{0.75\linewidth}{!}{%
\begin{tikzpicture}[x=1cm,y=1cm,font=\normalfont\small,line width=.55pt]
    \definecolor{mapblue}{HTML}{2761A3}

    \begin{scope}
        \coordinate (Y1) at (0,0);
        \coordinate (Y2) at (6,0);
        \coordinate (Y3) at (3,5.19615242);
        \coordinate (M) at (3,0);
        \coordinate (N) at (1.5,2.59807621);
        \coordinate (Q1) at (4,1.73205081);
        \coordinate (Q2) at (1.5,.86602540);
        \coordinate (Q3) at (2.5,2.59807621);
        \coordinate (E1) at (1.5,0);
        \coordinate (E2) at (4.5,0);
        \coordinate (E3) at (4.5,2.59807621);
        \coordinate (E4) at (3,2.59807621);
        \coordinate (E5) at (2.25,1.29903811);
        \coordinate (E6) at (.75,1.29903811);
        \coordinate (E7) at (2.25,3.89711432);

        \fill[mapblue!20] (M)--(E4)--(Q1)--cycle;
        \foreach \v in {Y2,Y3,M,E3,E4,E2}
            \draw[line width=.35pt] (Q1)--(\v);
        \foreach \v in {Y1,M,N,E1,E5,E6}
            \draw[line width=.35pt] (Q2)--(\v);
        \foreach \v in {Y3,N,M,E7,E5,E4}
            \draw[line width=.35pt] (Q3)--(\v);
        \draw (Y1)--(Y2)--(Y3)--cycle;
        \draw (Y3)--(M) (N)--(M);
        \draw[line width=.7pt] (M)--(E4)--(Q1)--cycle;

        \foreach \v in {Y1,Y2,Y3,M,N,Q1,Q2,Q3,E1,E2,E3,E4,E5,E6,E7}
            \fill (\v) circle (1.2pt);
        \foreach \v in {M,E4,Q1}
            \fill (\v) circle (2pt);
        \node[below=4pt] at (Y1) {$y_1$};
        \node[below=4pt] at (Y2) {$y_2$};
        \node[above=4pt] at (Y3) {$y_3$};
        \node[below=4pt] at (M) {$q_0$};
        \node at (3.25,2.95) {$q_1$};
        \node at (4.45,1.85) {$q_2$};
        \node at (3,5.95) {$\Delta_3$};
    \end{scope}

    \draw[->,line width=.65pt] (6.55,2.7)--(7.45,2.7)
        node[midway,above=4pt] {$R_1$};

    \begin{scope}[xshift=8cm]
        \coordinate (X1) at (0,0);
        \coordinate (X2) at (6,0);
        \coordinate (X3) at (3,5.19615242);
        \coordinate (M12) at (3,0);
        \coordinate (M13) at (1.5,2.59807621);
        \coordinate (M23) at (4.5,2.59807621);
        \coordinate (C) at (3,1.73205081);

        \fill[mapblue!20] (C)--(M13)--(X1)--cycle;
        \foreach \v in {X1,X2,X3,M12,M13,M23}
            \draw[line width=.35pt] (C)--(\v);
        \draw (X1)--(X2)--(X3)--cycle;
        \draw[line width=.7pt] (C)--(M13)--(X1)--cycle;

        \foreach \v in {X1,X2,X3,M12,M13,M23,C}
            \fill (\v) circle (1.2pt);
        \foreach \v in {C,M13,X1}
            \fill (\v) circle (2pt);
        \node[below=4pt] at (X1) {$r_2=x_1$};
        \node[below=4pt] at (X2) {$x_2$};
        \node[above=4pt] at (X3) {$x_3$};
        \node[left=4pt] at (M13) {$r_1$};
        \node at (3.55,2.35) {$r_0$};
        \node at (3,5.95) {$\Delta_3$};
    \end{scope}
\end{tikzpicture}%
}
\caption{A simplicial map $R_1:\Delta_3\to\Delta_3$.
The shaded simplices correspond under $q_i\mapsto r_i$.}
\label{fig:simplicial-map}
\end{center}

\end{example}

Additionally, we need some basic properties of product spaces from \cite[Theorem 18.2, Theorem 18.4 and Theorem 26.7]{munkres2000topology}.
\begin{proposition}
    \label{prop:product space}
    The following hold:
    \begin{enumerate}
    \item Let $X \subseteq \RR^m$ and $Y \subseteq \RR^n$ be polytopes.
    The function $f \colon X \times Y \to X \times Y$ is continuous if and 
    only if the coordinate functions $f_1 \colon X \times Y \to X$ and 
    $f_2 \colon X \times Y \to Y$ are continuous.
    \item Let $f\colon X \to \RR^r$ be continuous. Then the function
    $f \circ \pi_X\colon X \times Y \to \RR^r$ is continuous.
    \item For two simplices $\Delta_m$ and $\Delta_n$, the product
    $\Delta_m \times \Delta_n$ is a nonempty compact convex subset of
    $\RR^{m+n}$.
    \end{enumerate}
\end{proposition}    

Next, we introduce related complexity classes and tools needed.
A polynomial-time search reduction consists of
polynomial-time computable maps $f$ and $g$ such that,
for every instance $I$ of the source problem and every
solution $z$ of the target instance $f(I)$, the output
$g(I,z)$ is a solution of $I$.
The class $\PPAD$ consists of total search problems
reducible to \textsc{End-of-Line}. Given
an implicitly represented directed graph with indegree
and outdegree at most one and a specified source,
find another endpoint~\cite{papadimitriou1994complexity}.

We apply the general class defined by Etessami and Yannakakis
~\cite{etessami2010complexity} for fixed-point problems
with piecewise-linear, polynomial-time computable functions,
without constructing the underlying \textsc{End-of-Line}
graph.
Here, piecewise linearity is understood with respect to
a suitable polyhedral subdivision, which need not be
compatible with the simplicial complexes used in our
construction.

\begin{definition}[Polynomial piecewise-linear fixed-point problems]
    \label{def:ey-polynomial-pl}
    For each instance $I$, let
    \[
        F_I:D_I\to D_I
    \]
    be a continuous self-map of a nonempty compact convex
    domain $D_I$.
    This defines a
    \emph{polynomial piecewise-linear fixed-point problem}
    if the following hold:
    \begin{enumerate}
        \item 
        $D_I$ is a polytope defined by a set of linear
        inequalities.
        And coefficients can be
        computed from $I$ in polynomial time.

        \item
        There is a collection of hyperplanes dividing
        $D_I$ into polyhedral regions on each of which $F_I$ is affine.
        For each cell, the coefficients of its affine formula,
        as well as those of each dividing hyperplane, are rational
        numbers with encoding lengths polynomial in $|I|$ uniformly.

        \item
        There is a polynomial-time algorithm that
        interacts with a linear-comparison oracle for the point $x$.
        Each query asks whether $a^\top x\leq b$ with rational input $a$ and $b$, 
        and may depend on the answers to previous queries.
        If $x\notin D_I$, the algorithm outputs \textsc{No};
        otherwise, it outputs a rational matrix $C$ and a
        rational vector $c$ specifying the local affine formula
        applicable at $x$, so that $F_I(x)=Cx+c$.
    \end{enumerate}
\end{definition}

The following exact fixed-point theorem is due to
Etessami and Yannakakis~\cite[Theorem~5.2]{etessami2010complexity}.

\begin{theorem}
    \label{thm:ey-exact-fixed-point}
    For every instance $I$ of a polynomial piecewise-linear
    fixed-point problem, there exists a rational fixed point $x$
    s.t. $F_I(x) = x$ and its encoding length is polynomial in $|I|$.
    The search problem of computing such an exact
    rational fixed point belongs to $\PPAD$.
\end{theorem}

Thus, to apply \Cref{thm:ey-exact-fixed-point}, we will
construct a continuous self-map satisfying
\Cref{def:ey-polynomial-pl} and give a polynomial-time
procedure that transforms each polynomial-size rational
fixed point into a quasi-strict Nash equilibrium.

Our existence proof of quasi-strict equilibrium uses radial projection in a similar way to the existence proof of Nash~\cite{nash1951noncooperative} by Brouwer's fixed point theorem, which does not give a piecewise linear self-map. We thus replace the radial projection by Euclidean projection to obtain this, which may be computed as follows.
This may be computed in a simple way as shown in by a piecewise linear algebraic circuit as shown in the proof of \cite[Lemma 4.6]{etessami2010complexity}, and adapted in the following lemma.
\begin{lemma}[Euclidean projection]
    \label{lem:ey-modification}
    For $z\in \RR^k$, let
    $z_{(1)}\geq\cdots\geq z_{(k)}$ be its coordinates
    in decreasing order, and define
    \[
        t_k(z)
        =
        \max_{1\leq a\leq k}
        \frac{\sum_{i=1}^{a}z_{(i)}-1}{a},
    \] and for each $i \in [k]$,
    \[
        z_i \mapsto \max\{z_i-t_k(z),0\}.
    \]
    Then $t_k(z)$ is the unique real number $t$
    satisfying
    \[
        \sum_{i=1}^k\max\{z_i-t,0\}=1.
    \]
    This projection maps $\RR^k$ into $\Delta_k$, and is continuous and piecewise linear. Moreover, there is a polynomial-time algorithm
    with access to a linear-comparison oracle that outputs
    the local affine formula with input $z \in \RR^k$.
\end{lemma}

The complexity class used in our result about multiplayer games is 
the class $\exists \RR$. Its canonical complete problem,
the existential theory of the reals (ETR), asks whether a sentence of the form
\[
\exists x_1, \ldots, x_n \in \RR: \varphi(x_1, \ldots, x_n)
\]
is true, where $\varphi$ is a quantifier-free formula with equalities and inequalities
involving polynomials with rational coefficients. One may think of $\exists\RR$ as an
analogue of $\np$, and ETR as an analogue of $\SAT$. It is easy to see that
$\np \subseteq \exists\RR$. Canny proved that $\exists\RR \subseteq \pspace$ in \cite{canny1988some}.\footnote{A recent preprint OpenAI-141~\cite{OAI:Existential-universal-real-sentences-in-the-counting-hierarchy-October-4-2026}
claims the stronger upper bound $\exists\RR \subseteq \ch$, where $\ch$ denotes
the counting hierarchy.}

We use the minmax hardness theorem in 3-player games given by Hansen~\cite[Theorem~1]{hansen2019real} to establish $\exists\RR$-hardness. We use the specific formulation given in \cite[Definition~1 and Proposition~3]{berthelsen2022computational} and restated below (without the additional promise that one may assume $x_i = y_i \leq 1/2$ in the case of the minmax value being~0).
\begin{theorem}
    \label{thm:promised-minmax}
    Let $G$ be a three-player game $G$ with pure strategy sets
    $A,X,Y$ and a rational valued payoff function $r$ of player~1. Suppose
    $A=\{a_k^+,a_k^-:k\in[t]\}$ and
    \[
        r(a_k^+,b,c)=-r(a_k^-,b,c)
        \qquad \forall k\in[t],\ b\in X,\ c\in Y.
    \]
    Define
    \[
        V(G)=\min_{x\in\Delta_{X},\,y\in\Delta_{Y}}
                    \max_{a\in A}r(a,x,y).
    \]
    It is then $\exists\RR$-hard to decide whether $V(G)=0$ under
    the promise that the game $G$ meets one of the following conditions:
    \begin{enumerate}
        \item $V(G)>0$;
        \item $V(G)=0$, and there exist full-support strategies
        $x\in\Delta_X$ and $y\in\Delta_Y$ such that
        $r(a,x,y)=0$ for every $a\in A$.
    \end{enumerate}
\end{theorem}

\section{Existence of quasi-strict Nash equilibria in bimatrix games}

We use the notation introduced in the preliminary 
section. The best-response regions are
\[
\begin{aligned}
    Q_i
    &=\{y\in\Delta_n:i\in BR_1(y)\}
      =\{y\in\Delta_n:(Ay)_i\geq(Ay)_h
          \text{ for all }h\in M\},\\
    P_j
    &=\{x\in\Delta_m:j\in BR_2(x)\}
      =\{x\in\Delta_m:(B^\top x)_j\geq(B^\top x)_\ell
          \text{ for all }\ell\in N\}.
\end{aligned}
\]
Let $\mathcal C_n$ consist of all nonempty faces of the nonempty
$Q_i$, and let $\mathcal C_m$ consist of all nonempty faces of the
nonempty $P_j$. If
$P_j\cap P_\ell\neq\varnothing$, their intersection is obtained in
both regions by imposing the equality
$(B^\top x)_j=(B^\top x)_\ell$, so it is a face of both.
Consequently, $\mathcal C_m$ is a polyhedral complex with
$|\mathcal C_m|=\Delta_m$. The same argument gives
$|\mathcal C_n|=\Delta_n$.

For each $F\in\mathcal C_m$ and $G\in\mathcal C_n$, fix points
\begin{equation}
    \label{eq: int points}
    p_F\in\relint F,
    \qquad
    q_G\in\relint G.
\end{equation}
Using the construction from the preliminary section, set
\[
    \mathcal K_m=\mathcal K(\mathcal C_m,p),
    \qquad
    \mathcal K_n=\mathcal K(\mathcal C_n,q).
\]
By \Cref{prop:subdivision}, these are simplicial complexes and
\[
    |\mathcal K_m|=\Delta_m,
    \qquad
    |\mathcal K_n|=\Delta_n.
\]

For each nonempty $I\subseteq[k]$, define
\[
    \ell_k(I)=\frac{1}{|I|}\sum_{i\in I}e_i.
\]
This is the barycenter of the face $\conv\{e_i:i\in I\}$ of
$\Delta_k$. Let $\mathcal U_k$ be the barycentric subdivision of
$\Delta_k$. Its vertices are
the points $\ell_k(I)$, and a set of these vertices spans a simplex
exactly when their indexing sets form a chain under inclusion.
Moreover,
\[
    |\mathcal K_m|=|\mathcal U_m|=\Delta_m,
    \qquad
    |\mathcal K_n|=|\mathcal U_n|=\Delta_n.
\]

\begin{theorem}[Norde~\cite{norde1999bimatrix}]
    \label{thm:qsne-existence}
    Every bimatrix game has a quasi-strict Nash equilibrium.
\end{theorem}

\begin{proof}
    We construct four simplicial maps as follows:
    \begin{equation}
        \label{eq:four-vertex-maps}
        \begin{aligned}
            S_1(p_F)&=\ell_m(\supp(p_F)),&
            R_2(p_F)&=\ell_n(BR_2(p_F)),\\
            S_2(q_G)&=\ell_n(\supp(q_G)),&
            R_1(q_G)&=\ell_m(BR_1(q_G)).
        \end{aligned}
    \end{equation}
    All these maps send vertices of the simplicial 
    complex $\mathcal K_m$(or $\mathcal K_n$) to 
    vertices of the simplicial complex $\mathcal U_m$
    (or $\mathcal U_n$). 
    We verify that whenever vertices span a simplex
    of $\mathcal K$, their images span a simplex of
    $\mathcal U$. Then we use the same notation for 
    their continuous extensions:
    \[
    \begin{aligned}
    S_1&\colon\Delta_m\to\Delta_m,
    &R_2&\colon\Delta_m\to\Delta_n,\\
    S_2&\colon\Delta_n\to\Delta_n,
    &R_1&\colon\Delta_n\to\Delta_m.
    \end{aligned}
    \]
    
    If $F\subseteq F'$, then
    \begin{equation}
        \label{eq:vertex-monotonicity}
        \supp(p_F)\subseteq\supp(p_{F'}),
        \qquad
        BR_2(p_F)\supseteq BR_2(p_{F'}).
    \end{equation}
    
    To prove the first inclusion, suppose that
    $i\in\supp(p_F)$ but $i\notin\supp(p_{F'})$.
    Then $(p_F)_i>0$ and $(p_{F'})_i=0$.
    Since $p_{F'}\in\relint F'$, there exists $\delta>0$
    such that every point in the affine hull of $F'$
    at distance less than $\delta$ from $p_{F'}$
    belongs to $F'$.
    Moreover, $p_F\in F\subseteq F'$, so the point
    $z_\varepsilon=p_{F'}+\varepsilon(p_{F'}-p_F)$
    lies in this affine hull.
    Choose $\varepsilon>0$ sufficiently small and then 
    $z_\varepsilon\in F'$.
    This contradicts $F'\subseteq\Delta_m$.
    Therefore $\supp(p_F)\subseteq\supp(p_{F'})$.

    For the second inclusion, take any $j\in BR_2(p_{F'})$.
    By the definition of $P_j$, we have $p_{F'}\in P_j$.
    Thus $P_j$ is nonempty and belongs to $\mathcal C_m$.
    Since $F'$ and $P_j$ belong to the polyhedral complex
    $\mathcal C_m$, their nonempty intersection $F'\cap P_j$
    is a face of $F'$.
    Moreover, $p_{F'}\in(F'\cap P_j)\cap\relint F'$.
    A proper face of $F'$ cannot contain a point of
    $\relint F'$. We obtain $F'\cap P_j=F'$ and then $F'\subseteq P_j$.
    It implies that
    $p_F\in F\subseteq F'\subseteq P_j$ and $j\in BR_2(p_F)$.
    
    These arguments are identical for $G \subseteq G'$,
    \begin{equation}
    \label{eq:vertex-monotonicity-2}
    \supp(q_G)\subseteq\supp(q_{G'}),
    \qquad
    BR_1(q_G)\supseteq BR_1(q_{G'}).
    \end{equation}

    Let $\sigma\in\mathcal K_m$.
    By the definition of $\mathcal K_m$, there is a face chain
    \[
    F_0\subsetneq\cdots\subsetneq F_r
    \]
    in $\mathcal C_m$ such that
    \[
    \sigma=\conv\{p_{F_0},\ldots,p_{F_r}\}.
    \]
    Set $I_t=\supp(p_{F_t})$.
    By \Cref{eq:vertex-monotonicity},
    \[
    I_0\subseteq\cdots\subseteq I_r.
    \]
    Hence the corresponding faces of $\Delta_m$ satisfy
    \[
    \conv\{e_i:i\in I_0\}
    \subseteq\cdots\subseteq
    \conv\{e_i:i\in I_r\}.
    \]
    Their barycenters are precisely
    \[
    \ell_m(I_t)=S_1(p_{F_t}).
    \]
    By the definition of $\mathcal U_m$, these barycenters
    therefore span a simplex:
    \[
    \conv\{S_1(p_{F_0}),\ldots,S_1(p_{F_r})\}
    \in\mathcal U_m.
    \]

    Similarly, we have
    \[
    BR_2(p_{F_r})\subseteq\cdots\subseteq BR_2(p_{F_0}).
    \]
    The corresponding faces of $\Delta_n$ form a face chain,
    whose barycenters are $R_2(p_{F_t})$. Thus
    \[
    \conv\{R_2(p_{F_0}),\ldots,R_2(p_{F_r})\}
    \in\mathcal U_n.
    \]
    The same argument applies to $S_2$ and $R_1$
    on each simplex of $\mathcal K_n$.
    By \Cref{prop:simplicial map}, the four vertex
    maps extend to continuous simplicial maps.
    
    Following Nash's modification in the proof of
    \cite{nash1951noncooperative},
    define
    \[
    \begin{aligned}
    \varphi_{1i}(x,y)
    &=\max\{0,(R_1(y))_i-(S_1(x))_i\},
    &&i\in M,\\
    \varphi_{2j}(x,y)
    &=\max\{0,(R_2(x))_j-(S_2(y))_j\},
    &&j\in N.
    \end{aligned}
    \]
    For each $(x,y)\in\Delta_m\times\Delta_n$,
    define the modified strategies
    \begin{equation}
    \label{eq:brouwer-map}
        x'=
        \frac{x+\sum_{i\in M}\varphi_{1i}(x,y)e_i}
             {1+\sum_{i\in M}\varphi_{1i}(x,y)}, \qquad
        y'=
        \frac{y+\sum_{j\in N}\varphi_{2j}(x,y)e_j}
             {1+\sum_{j\in N}\varphi_{2j}(x,y)}.
    \end{equation}
    Both vectors have nonnegative coordinates summing
    to one. Thus these formulas define a map
    \[
    T:\Delta_m\times\Delta_n
      \longrightarrow\Delta_m\times\Delta_n.
    \]
    The coordinate projections and the four simplicial
    maps are continuous by \Cref{prop:product space}, 
    so all the functions
    $\varphi_{1i}$ and $\varphi_{2j}$ are continuous.
    Hence $T$ is continuous.
    By \Cref{prop:product space},
    $\Delta_m\times\Delta_n$ is nonempty, compact, and convex.
    Since $T$ is a continuous self-map,
    Brouwer's fixed-point theorem gives a fixed point $(\bar x,\bar y)$.

    One can show    
    \begin{equation}
    \label{eq:identical labels}
    S_1(\bar x)=R_1(\bar y),
    \qquad
    S_2(\bar y)=R_2(\bar x).
    \end{equation}
    At the fixed point, the first equation in
    \Cref{eq:brouwer-map} gives
    \[
    \varphi_{1i}(\bar x,\bar y)
    =
    \bar x_i
    \sum_{h\in M}\varphi_{1h}(\bar x,\bar y).
    \]
    $\sum_{i\in\supp(\bar x)} (S_1(\bar x))_i = 1$,
    for $S_1(\bar x)$ vanishes outside $\supp(\bar x)$.
    Then we have
    \[
    \sum_{i\in\supp(\bar x)}
    \bigl((R_1(\bar y))_i-(S_1(\bar x))_i\bigr)
    =
    -\sum_{i\notin\supp(\bar x)}(R_1(\bar y))_i
    \leq 0.
    \]
    Consequently, there exists $i_0\in\supp(\bar x)$
    such that
    \[
    (R_1(\bar y))_{i_0}
    \leq (S_1(\bar x))_{i_0},
    \]
    and therefore $\varphi_{1i_0}(\bar x,\bar y)=0$.
    Since $\bar x_{i_0}>0$, the fixed-point equation yields
    \[
    \sum_{h\in M}\varphi_{1h}(\bar x,\bar y)=0.
    \]
    It follows that, for every $i \in M$,
    \[
    (R_1(\bar y))_i\leq(S_1(\bar x))_i.
    \]
    We obtain 
    \[
    R_1(\bar y)=S_1(\bar x)
    \] for $\sum_i (R_1(\bar y))_i = \sum_i (S_1(\bar x))_i$.
    The same argument applied to the second equation in \Cref{eq:brouwer-map} 
    gives $R_2(\bar x)=S_2(\bar y)$.

    The last step is to extract quasi-strict Nash equilibria from 
    the above fixed point~$(\bar x, \bar y)$.
    Choose the unique minimal simplices
    of $\mathcal K_m$ and $\mathcal K_n$ containing $\bar x$
    and $\bar y$, respectively. Each point lies in the relative
    interior of its chosen simplex. These simplices correspond
    to the face chains
    \[
        F_0\subseteq\cdots\subseteq F_r,
        \qquad
        G_0\subseteq\cdots\subseteq G_s.
    \]
    We can write the points in barycentric coordinates
    \[
        \bar x=\sum_{t=0}^r\lambda_t p_{F_t},
        \qquad
        \bar y=\sum_{u=0}^s\mu_u q_{G_u},
    \]
    where $\lambda_t>0$ for $t=0,\ldots,r$ and $\mu_u>0$
    for $u=0,\ldots,s$, with $\sum_{t=0}^{r}\lambda_t=1$
    and $\sum_{u=0}^{s}\mu_u=1$.
    Rewrite \Cref{eq:identical labels} with \Cref{eq:four-vertex-maps},
    \begin{align}
        \label{eq:index set 1}
        S_1(\bar x) = \sum_{t=0}^r\lambda_t\ell_m(\supp(p_{F_t})) =
        \sum_{u=0}^s\mu_u\ell_m(BR_1(q_{G_u})) = R_1(\bar y),\\
        \label{eq:index set 2}
        S_2(\bar y) = \sum_{u=0}^s\mu_u\ell_n(\supp(q_{G_u}))=
        \sum_{t=0}^r\lambda_t\ell_n(BR_2(p_{F_t})) = R_2(\bar x).
    \end{align}

    We claim that $(p_{F_r}, q_{G_0})$ is a quasi-strict Nash equilibrium.
    By \Cref{def:quasi-strict nash eq}, verify the conditions
    \[
        \supp(p_{F_r}) = BR_1(q_{G_0}),\qquad \supp(q_{G_0}) = BR_2(p_{F_r}).
    \]

    The first condition holds by \Cref{eq:vertex-monotonicity} and 
    \Cref{eq:vertex-monotonicity-2}. $\supp(p_{F_r})$ and $BR_1(q_{G_0})$ are the largest 
    index sets, respectively. It follows that they are identical as \Cref{eq:index set 1} holds.

    To verify the second condition, notice that
    \Cref{eq:index set 2} implies that $S_2(\bar y)$ and
    $R_2(\bar x)$ have the same set of maximizing indices,
    \[
        \argmax_{i}(S_2(\bar y))_i
        =
        \argmax_{i}(R_2(\bar x))_i.
    \]
    These sets are precisely $\supp(q_{G_0})$ and
    $BR_2(p_{F_r})$, respectively, since they are the
    smallest sets in the corresponding chains,
    each $\ell_n(I)$ is uniform on $I$, and all weights
    are strictly positive. Hence
    \[
        \supp(q_{G_0})=BR_2(p_{F_r}).
    \]
    
    Thus $(p_{F_r},q_{G_0})$ is a quasi-strict Nash equilibrium.
\end{proof}

\section{Complexity of quasi-strict Nash equilibria in bimatrix games}
We now make the construction in \Cref{thm:qsne-existence}
effective. Recall the four simplicial maps
$S_1,S_2,R_1,R_2$ from \Cref{eq:four-vertex-maps}
and replace Nash modification by the EY
modification in \Cref{lem:ey-modification}. Denote the EY modification by $\Pi_k$.
For $(x,y)\in \Delta_m\times\Delta_n$, set
\begin{equation}
\label{eq:function h}
    h_1(x,y)=x+R_1(y)-S_1(x),
    \qquad
    h_2(x,y)=y+R_2(x)-S_2(y).
\end{equation}
Define $T'(x,y)=(x',y')$ by
\begin{equation*}
    \begin{aligned}
        x'_i
        &=\max\{(h_1(x,y))_i-t_m(h_1(x,y)),0\},
        &&i\in M,\\
        y'_j
        &=\max\{(h_2(x,y))_j-t_n(h_2(x,y)),0\},
        &&j\in N.
    \end{aligned}
\end{equation*}
By \Cref{lem:ey-modification}, $T'(x,y) = (\Pi_m \circ h_1(x,y), \Pi_n \circ h_2(x,y))$ is a
continuous self-map of $\Delta_m \times \Delta_n$.

\begin{lemma}
    A fixed point $(x,y)$ of $T'$ implies
    $S_1(x)=R_1(y)$ and 
    $S_2(y)=R_2(x)$.
\end{lemma}

\begin{proof}
    First observe that
    \[
        \supp(S_1(x))\subseteq\supp(x),
        \qquad
        \supp(S_2(y))\subseteq\supp(y).
    \]
    Indeed, if $x_i=0$, every vertex with a positive
    coefficient in a barycentric representation of $x$
    has zero $i$th coordinate.
    By \Cref{eq:four-vertex-maps}, its image under $S_1$
    also has zero $i$th coordinate.
    Affine interpolation therefore gives
    $(S_1(x))_i=0$.
    The argument for $S_2$ is identical.

    Suppose that $(x,y)$ is a fixed point of $T'$.
    Since $R_1(y),S_1(x)\in\Delta_m$, the coordinates
    of $h_1(x,y)$ sum to one.
    The candidate $a=m$ in the definition of $t_m$
    is therefore zero, so $t_m(h_1(x,y))\geq 0$.

    For every $i\in \supp(x)$, the fixed-point equation gives
    \[
        x_i=(h_1(x,y))_i-t_m(h_1(x,y)),
    \]
    and hence
    \[
        (R_1(y))_i-(S_1(x))_i=t_m(h_1(x,y)).
    \]
    For $i \not \in \supp(x) $, the support property above gives
    $(S_1(x))_i=0$.
    Consequently,
    \[
        0 =\sum_{i\in [m]}
                \bigl((R_1(y))_i-(S_1(x))_i\bigr)
            =|\supp(x)| \cdot t_m(h_1(x,y))+\sum_{i\notin \supp(x)}(R_1(y))_i.
    \]
    Every term on the right is nonnegative.
    Since $\supp(x) \neq\varnothing$, it follows that $t_m(h_1(x,y))=0$
    and $(R_1(y))_i=0$ for all $i\notin I$.
    Thus $S_1(x)=R_1(y)$.
    Applying the same argument to the second component
    gives $S_2(y)=R_2(x)$.
\end{proof}

Recall \Cref{eq: int points}. In topology, a standard choice
of a relative-interior point for each face is its barycenter,
defined as the average of its vertices, as illustrated in \Cref{example: barycenter of best-response}. 
For our purposes, however, computing this average by enumerating the vertices
can be inefficient. A polytope in $\RR^n$ defined by $2n$
linear inequalities may have exponentially many vertices.
Moreover, the barycenter plays no special role in our
construction, so we instead use linear programming to
compute a relative-interior point.

\begin{lemma}
    \label{lem:face-interior-lp}
    For $F\in\mathcal C_m$, let
    \[
        I_F=\{i\in M:x_i>0\text{ for some }x\in F\},
        \qquad
        J_F=\{j\in N:F\subseteq P_j\}.
    \]
    Given these index sets, a rational point
    $p_F\in\relint F$ can be computed in polynomial time.
    The same holds for $G\in\mathcal C_n$.
\end{lemma}

\begin{proof}
    Select $j_0$ as the smallest index in the set $J_F$ and solve
    \[
    \begin{aligned}
        \text{maximize}\quad & \varepsilon \\
        \text{subject to}\quad
            &\sum_{i\in M}x_i=1,\\
            &x_i=0 && \forall i\in M\setminus I_F,\\
            &x_i\ge \varepsilon &&\forall i\in I_F,\\
            &(B^\top x)_j=(B^\top x)_{j_0}
                &&\forall j\in J_F,\\
            &(B^\top x)_{j_0}-(B^\top x)_j\ge \varepsilon
                &&\forall j\in N\setminus J_F.
    \end{aligned}
    \]
    Since $F$ is a face of $P_{j_0}$, the zero-coordinate
    and payoff-tie equalities recorded by $I_F$ and $J_F$
    cut out $F$ within $P_{j_0}$. The remaining inequalities are strict
    on $\relint F$, so the optimal value satisfies $\varepsilon^*>0$.
    Consequently, the $x$-component of every optimal
    solution lies in $\relint F$.
    Using a fixed deterministic LP algorithm~\cite{khachiyan1980polynomial,karmarkar1984new}
    selects the same point whenever the same face occurs,
    with polynomial running time and output encoding length.
    The construction for $G\in\mathcal C_n$ is analogous,
    using $Ay$ in place of $B^\top x$.
\end{proof}

To obtain our membership result, we verify the three conditions in
\Cref{def:ey-polynomial-pl}. The first is immediate. For the second condition, start with the property of function composition.

\begin{lemma}
    \label{lem:pl-composition}
    For each input instance $I$, let $f$ and $g$ be
    composable continuous maps, with all dimensions
    polynomially bounded in $|I|$.
    If both maps satisfy the
    second condition of \Cref{def:ey-polynomial-pl},
    then their composition satisfies the second condition
    as well.
\end{lemma}

\begin{proof}
    On a cell of $f$, write $f(x)=Ax+b$.
    Substituting $y=Ax+b$ into a dividing hyperplane
    $a^\top y=t$ for $g$ gives
    \[
        (a^\top A)x=t-a^\top b.
    \]
    If $a^\top A=0$, this equation introduces no new
    dividing hyperplane.

    Take all original dividing hyperplanes for $f$,
    together with all nontrivial hyperplanes obtained
    by above substitutions over all cells of $f$.
    This finite collection divides the domain of $f$
    into polyhedral regions on each of which a fixed
    affine formula for $f$ applies and its image lies
    in a region with a fixed affine formula for $g$.
    If these formulas are $Ax+b$ and $By+c$, respectively,
    then
    \[
        (g\circ f)(x)=BAx+Bb+c.
    \]

    Each new coefficient is a sum of at most the
    intermediate dimension plus one terms, each of
    which is a product of at most two original coefficients.
    Since the intermediate dimension and the encoding
    lengths of the original coefficients are polynomially
    bounded in $|I|$, so are the encoding lengths of all
    new coefficients.
    Finally, $g\circ f$ is continuous because $f$ and $g$
    are continuous. Since it is affine on each of the
    polyhedral regions constructed above, it is piecewise linear.
\end{proof}

We now verify the second condition in
\Cref{def:ey-polynomial-pl}. Recall the function we used in
\Cref{eq:function h}
\[
T'(x,y) = (\Pi_m \circ h_1(x,y), \Pi_n \circ h_2(x,y)).
\]
Let $I$ denote the input game.
Choose the points $p_F$ and $q_G$ for every $F \in \mathcal C_m$ and $G \in \mathcal C_n$ using
\Cref{lem:face-interior-lp}. It admits that their rational coordinates
have a polynomial encoding length of $|I|$. 
Select any $(m-1)$-simplex
$\sigma=\conv\{p_0,\ldots,p_{m-1}\}$ of $\mathcal K_m$, and
set the matrix $P_\sigma=[p_0\ \cdots\ p_{m-1}]$.
This matrix is invertible, for these points are affinely independent and $\mathbf1^\top p_i = 1$ for each valid $i$.
Let the matrices $U_\sigma$ and $V_\sigma$ contain the corresponding
vertex images under $S_1$ and $R_2$, respectively.
Formally, 
\[
U_\sigma = [\ell_m(\supp(p_0))\ \cdots\ \ell_m(\supp(p_{m-1}))],
\qquad
V_\sigma = [\ell_n(BR_2(p_0))\ \cdots\ \ell_n(BR_2(p_{m-1}))].
\]

Then, the local formulas are
\begin{equation}
\label{eq:quasi local formula}
    S_1(x)=U_\sigma P_\sigma^{-1}x,
    \qquad
    R_2(x)=V_\sigma P_\sigma^{-1}x.
\end{equation}
The entries of $U_\sigma$ and $V_\sigma$ are either zero
or reciprocals of positive integers at most $\max\{m,n\}$. Since the entries of $P_\sigma$ have polynomial encoding
length, standard bounds for rational matrix inversion
imply the same for $P_\sigma^{-1}$.
The same argument applies to $S_2$ and $R_1$ on a $(n-1)$-simplex $\eta \in \mathcal K_n$. Therefore, the coefficients of local formulas $h_1(x,y)$ and $h_2(x,y)$ have polynomial size.

Each $(m-1)$-simplex $\sigma$ and
$(n-1)$-simplex $\eta$ can be described by
\[
    \sigma=\left\{
        x\in\Delta_m:
        P_\sigma^{-1}x\ge0
    \right\},
    \qquad
    \eta=\left\{
        y\in\Delta_n:
        Q_\eta^{-1}y\ge0
    \right\}.    
\]
The union of dividing hyperplanes in above two sets represents the dividing hyperplanes of the cell $\sigma \times \eta$ in the space $\Delta_m \times \Delta_n$.
These coefficients of dividing hyperplanes have polynomial size as well.
Thus, both maps $h_1$ and $h_2$ meet the second condition in \Cref{def:ey-polynomial-pl}.

For $\Pi_m$, we index the affine pieces by the coordinates
retained after thresholding.
For an input $z$, let
$T=\{i\in[m]:z_i>t_m(z)\}$.
This set is nonempty, and the threshold equation gives
\[
    1=\sum_{i\in T}(z_i-t_m(z)),
    \qquad
    t_m(z)=\frac{\sum_{i\in T}z_i-1}{|T|}
    =:t_T(z).
\]
Thus fixing $T$ fixes an affine expression for the threshold.

For each nonempty $T\subseteq[m]$, consider the region
\[
    R_T=\left\{z\in\mathbb R^m:
        \begin{array}{ll}
            z_i\ge t_T(z),&i\in T,\\
            z_i\le t_T(z),&i\notin T
        \end{array}
    \right\}.
\]
On this region, the vector
\[
    (\Pi_m(z))_i=
    \begin{cases}
        z_i-t_T(z),&i\in T,\\
        0,&i\notin T
    \end{cases}
\]
is nonnegative and sums to one, so it is indeed the output
of $\Pi_m$ by uniqueness of the threshold.
Consequently, $\Pi_m$ is affine on $R_T$, including its
boundary, and these regions cover $\mathbb R^m$.

The required dividing hyperplanes are obtained from the
nontrivial equations $z_i=t_T(z)$ over all such $T$.
Their coefficients and those of the affine formulas have
encoding length $O(\log(m+1))$, hence polynomial in $|I|$.
Thus $\Pi_m$ satisfies the second condition of
\Cref{def:ey-polynomial-pl}.
The argument for $\Pi_n$ is identical.

Using \Cref{lem:pl-composition}, the composition maps $\Pi_m \circ h_1$
and $\Pi_n \circ h_2$ also meet the 
second condition in \Cref{def:ey-polynomial-pl}. 
Taking the union of the dividing hyperplane collections
for the two composition maps gives regions on which both
affine formulas apply. Concatenating these formulas
also satisfies the second condition.

Next is to locate simplices of 
$\mathcal K_m$ and $\mathcal K_n$ cotaining $x$ and $y$ respectively, using only linear comparisons in the original input. Once these simplices are known, the required affine formula follows from \Cref{eq:quasi local formula} and 
\Cref{lem:ey-modification}.

\begin{lemma}
    \label{lem:oracle-computation}
    Given a rational bimatrix game $I$,
    there exists a polynomial-time algorithm with access to 
    a linear-comparison oracle for $z\in\RR^{m+n}$
    that outputs \textsc{No} if $z\notin\Delta_m\times\Delta_n$.
    Otherwise, it returns the local formula $T'(z)$.
\end{lemma}

\begin{proof}
    Write $z = (x,y)$ and consider the two
    parts separately by setting the coefficients corresponding to the other one to zero in each query.
    Test the constraints of $\Delta_m \times \Delta_n$. If any test fails, reject.
    Recall the best-response regions are
    \[
    \begin{aligned}
    Q_i
    &=\{y\in\Delta_n:i\in BR_1(y)\}
      =\{y\in\Delta_n:(Ay)_i\geq(Ay)_h
          \text{ for all }h\in M\},\\
    P_j
    &=\{x\in\Delta_m:j\in BR_2(x)\}
      =\{x\in\Delta_m:(B^\top x)_j\geq(B^\top x)_\ell
          \text{ for all }\ell\in N\}.
    \end{aligned}
    \]
    Use the oracle to locate a concrete pair of polytopes $(P, Q)$ s.t. $x \in P$ and $y \in Q$.
    Since each \(P_j\) has an explicit rational inequality description
    of polynomial encoding length, a strong optimization oracle for
    \(P_j\) can be implemented in polynomial time by rational linear
    programming. Hence, for every nonempty \(P_j\), its dimension can
    be computed in polynomial time by
    \cite[Lemma~6.5.3 and Remark~6.5.4(b)]{gls}.
    We can always choose the polytopes such that $\dim P = m-1$ and $\dim Q = n-1$
    since the full-dimensional regions $P_j$ must cover $\Delta_m$.
    Otherwise, the union of these closed sets would leave
    a full-dimensional open set in $\Delta_m$ covered only by finitely many
    lower-dimensional regions, which is impossible.
    The same holds for the regions $Q_i$ in $\Delta_n$.
    
    For the selected region $P$, first compute its label sets
    by linear programming:
    \[
    \begin{aligned}
        I_P
        &=\left\{a\in M:\max_{v\in P}v_a>0\right\},\\
        J_P
        &=\left\{b\in N:
        \max_{v\in P}
        \bigl((B^\top v)_j-(B^\top v)_b\bigr)=0
        \right\}.
    \end{aligned}
    \]
    These computations use only the explicit inequality description
    of $P$.
    Given these labels, apply \Cref{lem:face-interior-lp}
    to obtain the prescribed point $p\in\relint P$.
    The labels and interior point for $Q$ are computed analogously.

    Then we need to compute the corresponding $(m-1)$-simplex in $P$ and the corresponding $(n-1)$-simplex in $Q$, 
    respectively. Equivalently, compute two sets of points $\{p_0, \cdots, p_{m-1}\}$ and $\{q_0, \cdots, q_{n-1}\}$
    such that $x \in \conv\{p_0, \cdots, p_{m-1}\}$ and $y \in \conv\{q_0, \cdots, q_{n-1}\}$. 
    We next show how to compute $p_{m-1},\ldots,p_0$ recursively.
    Set $F_{m-1}=P$, $p_{m-1}=p$, and $x_{m-1}=x$.
    The initial label sets are
    $I_{m-1}=I_P$ and $J_{m-1} = J_P$.
    
    At stage $i$, we maintain a polytope $F_i$,
    its label sets $I_i=I_{F_i}$ and $J_i=J_{F_i}$,
    its prescribed relative-interior point $p_i$,
    and a matrix $C_i$ defining target point $x_i\in F_i$.
    In each stage, first
    test whether $x_i = p_i$. We can compare them directly using oracle for the first stage.
    As shown later, this test can also be achieved using linear comparisons in the original input $x$ for other stages.
    
    If $x_i\ne p_i$, consider the ray
    \[
    r_i(t)=p_i+t(x_i-p_i),\qquad t\ge0.
    \]
    Let $x_{i-1}$ be its first intersection with a facet of $F_i$,
    and choose that facet as $F_{i-1}$, breaking ties by a fixed ordering.

    To find this intersection, we first determine the facets of \(F_i\)
    from its label sets \(I_i\) and \(J_i\).
    Fix \(j_0=\min J_i\), and consider the inequalities
    \[
    z_a \ge 0 \quad \forall a\in I_i,
    \qquad
    (B^\top z)_{j_0}-(B^\top z)_b \ge 0
    \quad \forall b\in N\setminus J_i.
    \]
These are the inequalities in the label-based description of \(F_i\)
that are not identically tight on \(F_i\).
For each such inequality \(h(z)\ge 0\), form
\[
    G_h=F_i\cap\{z:h(z)=0\},
\]
and retain \(h\) if \(G_h\) is nonempty and
\(\dim G_h=\dim F_i-1\).
These tests can be performed in polynomial time by rational
linear programming.
Let \(\mathcal H_i\) be the resulting set of functions.
Every facet of \(F_i\) is represented by at least one function
in \(\mathcal H_i\).
Each \(h\in\mathcal H_i\) is homogeneous linear, is nonnegative
on \(F_i\), and satisfies \(h(p_i)>0\), since
\(p_i\in\operatorname{relint}F_i\).

    For $h\in\mathcal H_i$, if the corresponding facet intersects with the ray,
    substituting the ray into $h(r_i(t))=0$ gives
    \[
    h(r_i(t)) = h(p_i) + t(h(x_i) - h(p_i)) = 0.
    \]
    We have
    \[
    t_h = \frac{1}{1 - h(x_i)/h(p_i)}.
    \]
    Choosing another $g \in \mathcal H_i$, if it also intersects with the ray,
    we can compare
    \[
    t_h\le t_g
    \quad\Longleftrightarrow\quad
    \frac{h(x_i)}{h(p_i)}
    \le
    \frac{g(x_i)}{g(p_i)} < 1.
    \]
    Otherwise, if the ray also misses the supporting hyperplane
    $\{z:g(z)=0\}$, then $g(x_i)\geq g(p_i)$.
    It is also possible that the ray meets this hyperplane outside
    $F_i$, possibly outside $\Delta_m$, while $g(x_i)/g(p_i)<1$.
    In this case, before reaching that intersection at $t_g$, the ray
    must meet some facet with defining function $f\in\mathcal H_i$
    at a parameter $t_f<t_g$. By the same comparison,
    \[
        \frac{f(x_i)}{f(p_i)}<\frac{g(x_i)}{g(p_i)}<1.
    \]
    Hence such an intersection outside $F_i$ cannot yield the minimum ratio.
    Thus the first facet encountered is obtained by choosing
    $h_*\in\mathcal H_i$ that minimizes $h(x_i)/h(p_i)$,
    breaking ties by the fixed ordering.

    To implement these comparisons using only linear queries in the
    original input $x$, store the target points as
    \[
    u_i=C_i x, \quad w_i = \mathbf1^\top u_i \quad \text{and} \quad 
    x_i=\frac{u_i}{w_i},
    \]
    where $C_i$ is a rational matrix with polynomial size and $\mathbf 1^\top u_i > 0$. The vectors \(u_i\) and points \(x_i\) are used only as symbolic quantities. The algorithm explicitly stores the rational matrix \(C_i\), rather than the numerical values of \(u_i\), \(w_i\), or \(x_i\).
    Initially, $C_{m-1}=Id_m$.
    The facet comparison becomes
    \[
    \frac{h(x_i)}{h(p_i)}
    \le
    \frac{g(x_i)}{g(p_i)}
    \quad \Longleftrightarrow \quad
    g(p_i)h(\frac{u_i}{w_i})\le h(p_i)g(\frac{u_i}{w_i})
    \quad \Longleftrightarrow \quad
    g(p_i)h(C_i x)\le h(p_i)g(C_i x),
    \]
    which is a linear comparison in the original $x$.
    After selecting $h_*$, the intersection is exactly
    \begin{equation*}
    \begin{aligned}
    x_{i-1}&
    =p_i + \frac{1}{1 - h_*(x_i)/h_*(p_i)}(x_i-p_i) \\
    &=p_i + \frac{w_i}{w_i - h_*(u_i)/h_*(p_i)}(\frac{u_i}{w_i}-p_i)\\
    &=\frac{u_i - p_i(h_*(u_i)/h_*(p_i))}{w_i - (h_*(u_i)/h_*(p_i))}.
    \end{aligned}
    \end{equation*}

    We therefore preserve the representation by setting
    \[
    u_{i-1}=u_i - p_i(h_*(u_i)/h_*(p_i)),
    \quad
    w_{i-1}=w_i - (h_*(u_i)/h_*(p_i)),
    \quad
    F_{i-1}=F_i\cap\{z:h_*(z)=0\}.
    \]
    Moreover, $h_*(u_i)/h_*(p_i) =
    w_i h_*(x_i)/h_*(p_i)< w_i$ based on above analysis.
    Writing $h_*(z)=a^\top z$, update the matrix by
    \begin{equation*}
    \begin{aligned}
        u_{i-1}&=u_i - p_i(h_*(u_i)/h_*(p_i)) \\
        &= C_i x - p_i \frac{h_*(C_i x)}{h_*(p_i)} \\
        & = C_i x - p_i \frac{a^\top C_i x}{a^\top p_i} = (Id_m - \frac{p_ia^\top}{a^\top p_i})C_i x.
    \end{aligned}
    \end{equation*}
    Thus, $C_{i-1} = (Id_m - \frac{p_i a^\top}{a^\top p_i})C_i$, which is a rational matrix.
    Not hard to verify that $Id_m - \frac{p_ia^\top}{a^\top p_i}$ has polynomial size for each $i$ and thus $C_i$ has polynomial size for each $i$.
    Additionally, $\mathbf1^\top p_i = 1$ since $p_i$ is a point in $\Delta_m$. We obtain $\mathbf1^\top u_{i-1} = \mathbf1^\top u_i - \mathbf1^\top p_i(h_*(u_i)/h_*(p_i)) = w_i - h_*(u_i)/h_*(p_i) = w_{i - 1} > 0$.

    To update the labels of the selected facet,
    delete coordinates identical to zero in $F_{i-1}$   
    and add new best-response actions in $F_{i-1}$ by linear programming, 
    \[
    \begin{aligned}
    I_{i-1}
    &=
    \left\{
        a\in I_i:
        \max_{v\in F_{i-1}}v_a>0
    \right\},\\
    J_{i-1}
    &=
    J_i\cup
    \left\{
        b\in N\setminus J_i:
        \max_{v\in F_{i-1}}
        \bigl((B^\top v)_{j_0}-(B^\top v)_b\bigr)=0
    \right\}.
    \end{aligned}
    \]
    Using the updated labels, apply \Cref{lem:face-interior-lp}
    to compute the prescribed point
    $p_{i-1}\in\relint F_{i-1}$.

    If $x_i=p_i$, complete the remaining face chain by choosing
    facets according to the fixed ordering and applying the same
    label updates and interior-point computations.
    No further target points are needed, and all subsequent
    barycentric coefficients are zero.

    We now address the issue above. Determining whether
    $p_i=x_i$ using a linear-comparison oracle for the original input $x$
    is equivalent to determining whether
    $C_i x - w_i p_i = 0$, which just uses linear comparisons.

The resulting complete face chain
$F_0\subsetneq\cdots\subsetneq F_{m-1}$
determines an $(m-1)$-simplex
\[
    \sigma=\conv\{p_0,\ldots,p_{m-1}\}
\]
containing $x$. The correctness can be proved by induction since for each valid $i$ $x_{i} = (h_*(x_i)/h_*(p_i))p_{i} +  (1 - (h_*(x_i)/h_*(p_i)))x_{i-1}$. The same argument applies to $y$.

Based on the above analysis, the local formulas
of all four simplicial maps can be determined by 
\Cref{eq:quasi local formula}. Execute the Euclidean projection algorithm of
\Cref{lem:ey-modification} on our formula
\[
    H(x,y)=(h_1(x,y),h_2(x,y))
          =C\binom{x}{y}+c.
\]
Each comparison $a^\top H(x,y)\le b$ becomes
\[
    (a^\top C)\binom{x}{y}\le b-a^\top c,
\]which is also a linear comparison in the original input.
The resulting local formula for $T'$ is obtained
in polynomial time. This concludes our proof.
\end{proof}

\ppad*

\begin{proof}
    For hardness, use \PPAD-completeness of computing a bimatrix Nash equilibrium by \cite{chen2009settling}.

    For membership, using \Cref{thm:ey-exact-fixed-point}, we can obtain
    a fixed point $(\bar x,\bar y)$ of $T'$, which is rational and can be computed in \PPAD.
    Execute the algorithm in \Cref{lem:oracle-computation} on input $(\bar x, \bar y)$. Since the input
    is rational, the oracle is no longer needed and all linear comparisons 
    can be finished in polynomial time.
    As the intermediate result of the algorithm, we have $\bar x \in \sigma$ and $\bar y \in \eta$,
    together with $P^{-1}_\sigma$ and $Q^{-1}_\eta$ defined in \Cref{eq:quasi local formula}.
    Compute the barycentric coordinates as follows
    \[ \lambda = P^{-1}_\sigma \bar x, \qquad \mu = Q^{-1}_\eta \bar y. \]
    Select
    \[
    r = \max\{i : \lambda_i > 0\}, \qquad s = \min \{j: \mu_j > 0\}.
    \]
    As the proof of \Cref{thm:qsne-existence} shows,
    $(p_r, q_s)$ is a quasi-strict Nash equilibrium.
\end{proof}

\section{Complexity of deciding whether quasi-strict Nash equilibria exist}
\existr*

\begin{proof}
    Containment in $\exists\RR$ for general-sum games is straightforward.
    Namely, let $P_{i,a}$ denote the expected payoff if player~$i$ chooses action~$a$ from
    action set $A_i$.
    Introduce a real variable $v_i$ to denote the best-response payoff of player~$i$ and a vector of real variables $\nu_i$
    to denote the strategy. 
    We can construct an existential formula
    \[
    \begin{aligned}
    \exists &((\nu_i(a))_{i,a},(v_i)_i):\\
    &\bigwedge_i
    \biggl(
    \sum_{a\in A_i}\nu_i(a)=1
    \;\land\;
    \bigwedge_{a\in A_i}
    \bigl[
        \bigl(\nu_i(a)>0 \land P_{i,a}(\nu_{-i})=v_i\bigr)
        \lor
        \bigl(\nu_i(a)=0 \land P_{i,a}(\nu_{-i})<v_i\bigr)
    \bigr]
    \biggr).
    \end{aligned}
    \]
    It uses only polynomial equalities and inequalities.

    For hardness, take an instance $G$ of
    \Cref{thm:promised-minmax}. Construct a three-player game
    $\widehat G$ with action sets
    \[
        \widehat A_1=A\cup\{\bot_1\},\qquad
        \widehat A_2=X\cup\{\bot_2\},\qquad
        \widehat A_3=Y\cup\{\bot_3\}.
    \]
    Define its payoffs in the following form, where player~1 selects the matrix, player~2 selects
    the row, and player~3 selects the column.
    The first matrix applies to any $a\in A$, with
    $r=r(a,b,c)$ in its $X\times Y$ block and all other displayed
    payoff vectors are constant on their blocks. Every payoff vector sums to zero.
    \[
    \begingroup
    \renewcommand{\arraystretch}{1.35}
    \setlength{\arraycolsep}{8pt}
    \begin{array}{c}
        \begin{array}{c|c|c|}
            \multicolumn{1}{c}{}
                &\multicolumn{1}{c}{Y}
                &\multicolumn{1}{c}{\bot_3}\\ \cline{2-3}
            X &(r,-r,0)&(0,0,0)\\ \cline{2-3}
            \bot_2 &(0,0,0)&(-1,0,1)\\ \cline{2-3}
        \end{array}\\[4pt]
        a\in A
    \end{array}
    \qquad
    \begin{array}{c}
        \begin{array}{c|c|c|}
            \multicolumn{1}{c}{}
                &\multicolumn{1}{c}{Y}
                &\multicolumn{1}{c}{\bot_3}\\ \cline{2-3}
            X &(0,0,0)&(0,0,0)\\ \cline{2-3}
            \bot_2 &(1,-1,0)&(0,-1,1)\\ \cline{2-3}
        \end{array}\\[4pt]
        a \in \{\bot_1\}
    \end{array}
    \endgroup
    \]
    
    The construction adds one action per player and uses only
    the original payoff entries, their negatives, and $0,\pm1$, so its size and
    encoding length are polynomial in the input size.

    Suppose first that $V(G)=0$. Choose the full-support
    strategies $x,y$ supplied by the promise.
    Let $\nu_1$ be uniform on $\widehat A_1$, and set
    \[
        \nu_2=(x_1, \ldots, x_{|X|},0),\qquad
        \nu_3=(\tfrac12 y_1, \ldots, \tfrac12 y_{|Y|},\tfrac12),
    \]
    where the last coordinate corresponds to the additional
    action. It's easy to check that all three strategies are valid.
    Moreover,
    \[
        BR_1(\nu_{-1})=\widehat A_1,\qquad
        BR_2(\nu_{-2})=X,\qquad
        BR_3(\nu_{-3})=\widehat A_3.
    \]
    The first equation follows from $r(a,x,y) = 0$ for each $a \in A$.
    For the second equation, $\nu_1$ assigns equal probabilities
    to each pair $a_k^+,a_k^-$, so every original action of
    player~2 has expected payoff zero. The action $\bot_2$
    yields payoff $-\nu_1(\bot_1)<0$.
    The last equation holds because $\nu_2(\bot_2)=0$,
    so every action of player~3 has payoff zero.
    Thus $(\nu_1, \nu_2, \nu_3)$ is a quasi-strict Nash equilibrium of $\widehat G$, for
    \[
    BR_1(\nu_{-1})=\supp(\nu_1) ,\qquad
    BR_2(\nu_{-2})=\supp(\nu_2),\qquad
    BR_3(\nu_{-3})=\supp(\nu_3).
    \]

    Conversely, let $(\nu_1, \nu_2, \nu_3)$ be a quasi-strict Nash equilibrium
    of $\widehat G$.
    We first show that $\nu_2(\bot_2)=0$.
    If $\nu_2(\bot_2)>0$, player~3's action $\bot_3$ has
    positive payoff, while every original action has payoff zero.
    Hence player~3 plays $\bot_3$ with probability one.
    Player~1's action $\bot_1$ then yields zero, while every
    original action yields $-\nu_2(\bot_2)<0$, so player~1
    plays $\bot_1$ with probability one.
    But player~2's action $\bot_2$ now yields $-1$, while every
    original action yields zero, contradicting
    $\nu_2(\bot_2)>0$.

    Since $\nu_2(\bot_2)=0$, every action of player~3 has
    payoff zero. QSNE requires full support on $\widehat A_3$.
    Suppose, for a contradiction, that $\nu_1(\bot_1)=0$.
    Player~1's action $\bot_1$ yields payoff zero. QSNE therefore requires
    player~1's equilibrium payoff to be strictly positive.
    Since player~3's payoff is zero and the game is zero-sum,
    player~2's equilibrium payoff must be strictly negative.
    However, when $\nu_1(\bot_1)=0$, player~2 can obtain zero
    by switching to $\bot_2$, which is a valuable deviation.
    This contradicts the Nash equilibrium condition, so
    $\nu_1(\bot_1)>0$.

    Define two normalized vectors
    \[
        x = \left( \nu_2(c) \right)_{c \in X}, \qquad y = \left( \frac{\nu_3(c)}{\sum_{d \in Y} \nu_3(d)} \right)_{c \in Y}.
    \]
    Since $\bot_1$ is played with positive probability and
    yields payoff zero, for each pair $a^+, a^- \in A$, we have
    \[
        r(a^+,x,y)\le0, \qquad r(a^-,x,y) \leq 0.
    \]
    The condition $r(a^+,x,y) = - r(a^-, x,y)$ in \Cref{thm:promised-minmax}
    implies $r(a,x,y)=0$ for every $a\in A$ and then $V(G)=0$.
    We have therefore proved, on the promised instances,
    \[
        V(G)=0
        \quad\Longleftrightarrow\quad
        \widehat G\text{ admits a quasi-strict Nash equilibrium}.
    \]
    This establishes $\exists\RR$-hardness for three-player
    zero-sum games.

    Given any fixed $n>3$, add $n-3$ players, each with a single
    action and constant zero payoff, and leave the original
    payoffs independent of these players. This preserves the
    existence of quasi-strict Nash equilibria and the zero-sum
    property, completing the proof.
\end{proof}

\section*{AI methodology}
ChatGPT-6 Astra was used to assist in developing the proofs, and in particular provided a correction to the step of the proof that recovers a quasi-strict equilibrium from a fixed point. This tool was also used to assist in writing the paper and preparing the figures. The authors have verified all proofs and naturally take full responsibility for all mathematical claims given in the paper.

\section*{Acknowledgements}
Kristoffer Arnsfelt Hansen and Xinhao Nie were supported by the Independent Research Fund Denmark under grant no.~10.46540/4283-00272B. This work was also supported in part by a grant of access to OpenAI models through the ChatGPT for Academic Researchers program.

\bibliographystyle{alpha}
\bibliography{bib}

\end{document}